\documentclass[aps,prb,reprint,amsmath,amssymb,superscriptaddress]{revtex4-2}

\usepackage{graphicx}
\usepackage{dcolumn}
\usepackage{bm}

\begin{document}

\title{Dynamical Signatures of Selective Gap Closing in Multi-Dimensional Monotone Quasiperiodic Operators}

\author{Leonid Parnovski}
\affiliation{Department of Mathematics, University College London, London WC1E 6BT, UK}
\author{Roman Shterenberg}
\affiliation{Department of Mathematics, University of Alabama at Birmingham, Birmingham, AL 35294, USA}

\date{\today}

\begin{abstract}
We translate the spectral gap-closing criterion for multi-dimensional monotone quasiperiodic Schr\"odinger operators into a concrete dynamical protocol and verify it by large-scale numerics. For a non-principal label we construct a sampling function whose leading eigenfunction weight at the origin vanishes; a local bump then drives the residual weight through zero. The height of a weighted oscillatory time average of the origin-to-localization-center correlator exhibits a deep minimum precisely at the closing, while the site-control label remains featureless and orders of magnitude larger. The signature is robust and maps directly onto superconducting-circuit and photonic platforms.
\end{abstract}

\maketitle

\section{Introduction}
Multi-dimensional quasiperiodic operators with monotone sawtooth potentials realize Anderson localization together with a Cantor spectrum of positive measure~\cite{Kachkovskiy2026}. In the weak-hopping regime a gap labeled by a lattice point $n$ is open if and only if the corresponding eigenfunction does not vanish at the origin. Principal labels open automatically for small hopping, while non-principal labels can be closed by an arbitrarily small $C^1$ perturbation of the sampling function. This distinction is invisible in one dimension and cannot be reduced to an algebraic sum of one-dimensional Cantor sets.

The same lattice Hamiltonians arise as tight-binding models for electrons in quasicrystals~\cite{Damanik2015}, cold atoms in optical lattices~\cite{Aidelsburger2013,Viebahn2019,Roati2008,Schreiber2015}, and on-chip architectures (superconducting resonator or transmon arrays~\cite{Li2023,deAlbornoz2023}, photonic-crystal cavities~\cite{Ye2022}, and, in a continuum limit, moir\'e superlattices~\cite{Dean2013,Hunt2013}). In each setting a spectral gap is a stop-band, while a vanishing eigenfunction weight at a distinguished site corresponds to a dark mode. The present work translates the spectral criterion of Ref.~\cite{Kachkovskiy2026} into a concrete dynamical protocol and verifies it numerically, thereby providing an experimentally relevant signature of selective gap closing.

\section{Model and parameters}
The Hamiltonian on $\ell^2(\mathbb{Z}^2)$ is
\begin{equation}
(H\psi)(n)=\varepsilon(\Delta\psi)(n)+\bigl[f(n\cdot\omega)+s\,g(n\cdot\omega)\bigr]\psi(n).
\end{equation}
We take the Diophantine frequencies $\omega_1\approx0.845751$, $\omega_2\approx0.376558$ and the piecewise-linear sampling function defined by the knots in Table~\ref{tab:knots}. These values are chosen so that the sum of the three length-3 path contributions to $\psi_{(1,2)}(0)$ vanishes identically, consistent with the first-order expansion developed in Ref.~\cite{Kachkovskiy2026}. Hopping $\varepsilon=0.01$. The bump $g$ is a smooth unit-height function supported near the phase $\omega_1$ corresponding to an intermediate site of the length-3 paths. To match the experimental setting, simulations use dynamical evolution on a centered grid of 99 sites ($[-5,5]\times[-3,5]$). Evolution time $T=200$; the quadratic Hann window is applied to the amplitude recorded at the origin after initialization at site $(1,2)$. The observable is the weighted oscillatory time average (Fourier peak) of the origin-to-localization-center correlator after initialization at the localization center. 

\begin{table}[ht]
\caption{Knots of the piecewise-linear sampling function $f$ that cancel the leading contribution to $\psi_{(1,2)}(0)$. Phases are taken modulo 1.}
\label{tab:knots}
\begin{ruledtabular}
\begin{tabular}{ccc}
Knot & $x$ (mod 1) & $f(x)$\\
\hline
0 & 0 & 0\\
$\omega_1+\omega_2$ & 0.2223 & 0.5\\
$\omega_2$ & 0.3766 & 0.7\\
$\omega_1+2\omega_2$ & 0.5989 & 1\\
$2\omega_2$ & 0.7531 & 1.6\\
$\omega_1$ & 0.8458 & 2.8\\
1 & 1 & 3\\
\end{tabular}
\end{ruledtabular}
\end{table}

\section{Numerical results}
\begin{figure}[ht]
\includegraphics[width=0.95\columnwidth]{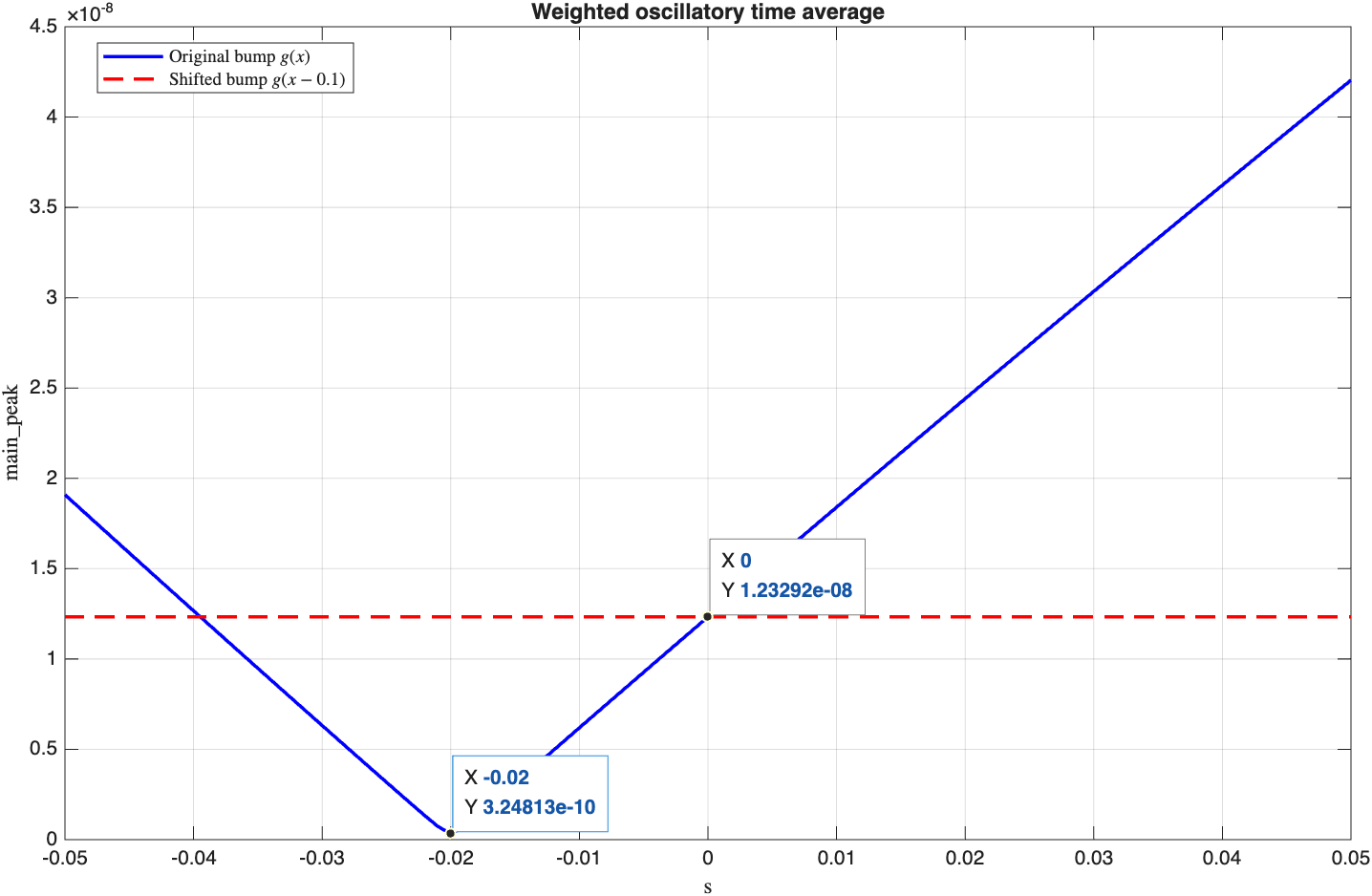}
\caption{Weighted oscillatory time average (main peak) versus bump amplitude $s$ for the label $(1,2)$. The original bump produces a clear minimum; the shifted-bump control is flat.}
\label{fig:main}
\end{figure}

Figure~\ref{fig:main} shows the weighted oscillatory time average of the correlator for the label $(1,2)$. With the original bump the peak exhibits a pronounced V-shaped minimum near $s\approx-0.02$, dropping to approximately $3.25\times10^{-10}$. The shifted-bump control remains featureless and roughly forty times larger. This contrast demonstrates that the dip is produced by the local perturbation of the cancelled weight and is not a global spectral shift, in accordance with the gap-closing criterion of Ref.~\cite{Kachkovskiy2026}. Because the leading length-3 contribution has already been cancelled by the choice of $f$, the residual weight is of higher order; the observed minimum is the first-order response of that residual to the bump.

\begin{figure}[ht]
\includegraphics[width=0.95\columnwidth]{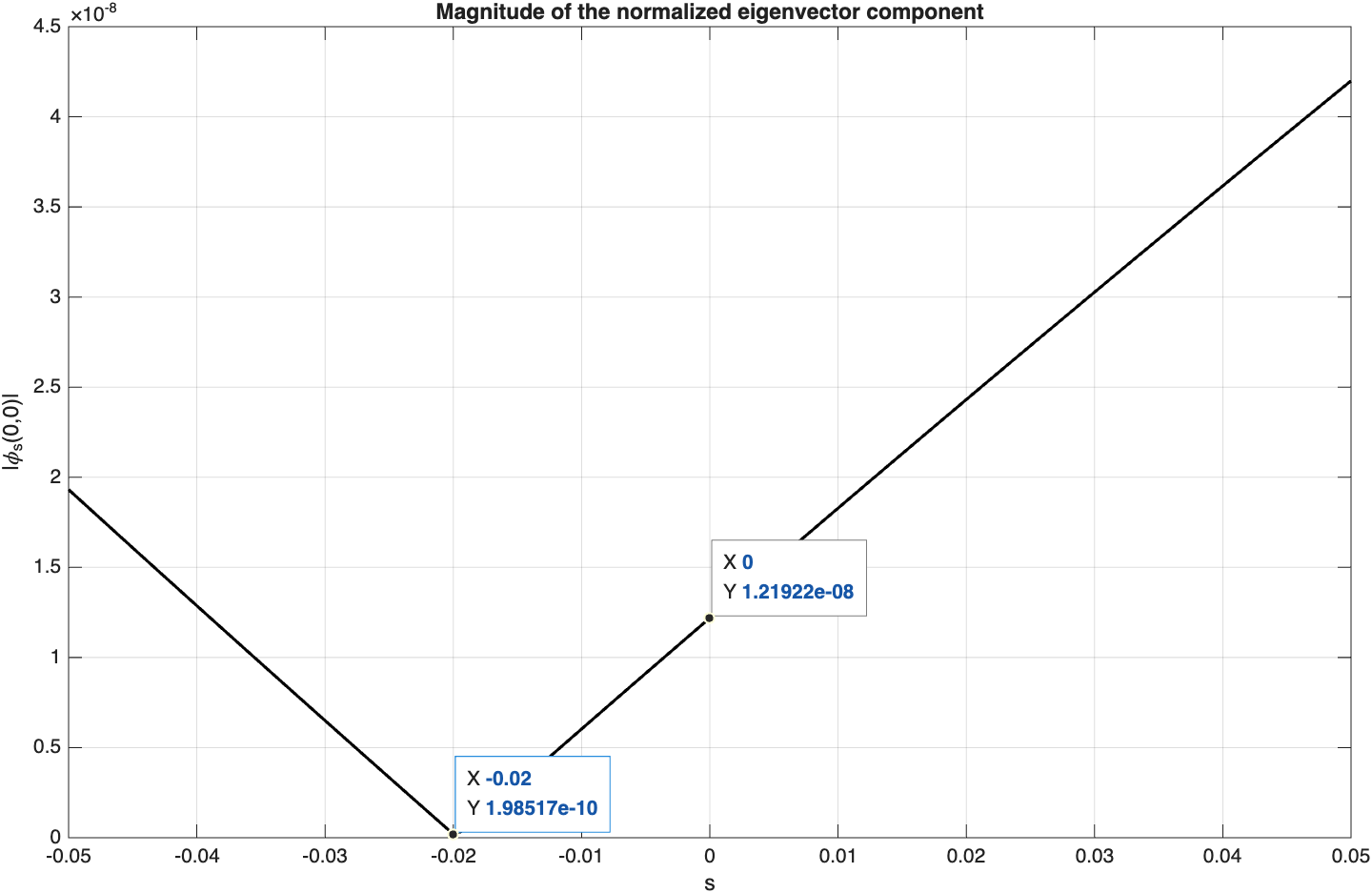}
\caption{Exact magnitude of the normalized eigenvector component $|\psi_{(1,2)}(0)|$ versus bump amplitude $s$. The minimum coincides with the dynamical dip in Fig.~\ref{fig:main}.}
\label{fig:exact}
\end{figure}

Figure~\ref{fig:exact} shows the exact magnitude of the normalized eigenvector component $|\psi_{(1,2)}(0)|$ obtained by direct diagonalization. It exhibits the same V-shaped minimum near $s\approx-0.02$, dropping to approximately $1.99\times10^{-10}$. The close agreement between the dynamical Fourier peak and the static eigenvector weight confirms that the weighted oscillatory average faithfully tracks the eigenfunction component that controls the gap.

\begin{figure}[ht]
\includegraphics[width=0.95\columnwidth]{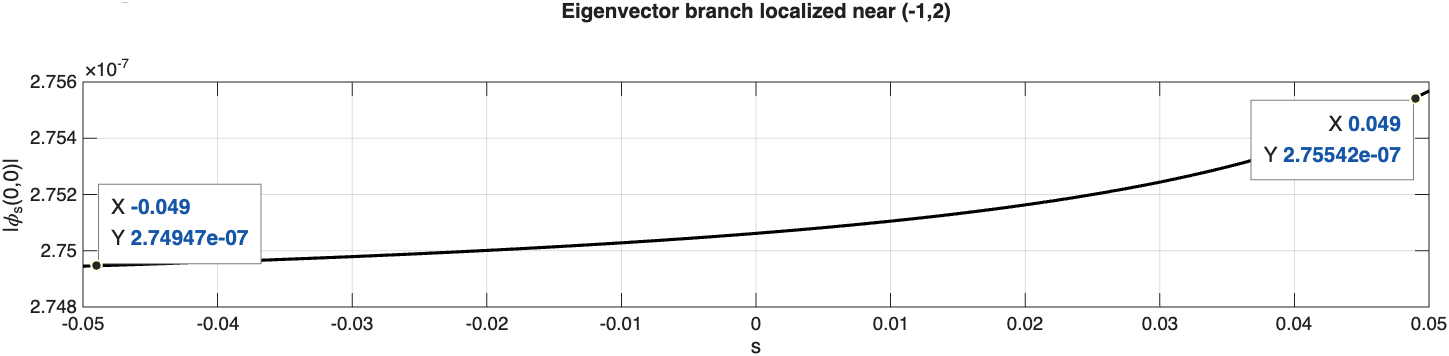}
\caption{Eigenvector weight at the origin for the branch localized near $(-1,2)$ versus bump amplitude $s$. The value stays near $2.75\times10^{-7}$ with no feature.}
\label{fig:site}
\end{figure}

Figure~\ref{fig:site} shows the eigenvector weight at the origin for the site-control label $(-1,2)$. This label lies at the same graph distance 3 from the origin as $(1,2)$. The weight remains nearly constant near $2.75\times10^{-7}$ across the scanned interval of $s$, with no dip. The contrast confirms that the leading contribution of the $(1,2)$ eigenfunction was cancelled by the special choice of $f$, while a nearby label of equal distance retains an unsuppressed spectral weight.

The hierarchy---a residual main peak that develops a deep minimum under the local bump, while the site-control weight remains large and featureless---is a bulk signature of the selective gap-closing mechanism. The same qualitative features (location and depth of the main dip, flatness of both controls, and absolute scale of each observable) remain unchanged (and nearly identical quantitatively) when the grid is varied over a wide range: from the modest rectangular domain $[-5,5]\times[-3,5]$ (99 sites) up to large square domains $[-50,50]\times[-50,50]$ (ten thousand sites). Increase of the evolution time $T$ naturally improves the sharpness of the dip. This stability demonstrates that the dynamical signature is insensitive to the precise infrared cutoff and to the precise length of the Fourier window, and is therefore compatible with the finite coherence times and finite array sizes available in present-day superconducting and photonic experiments~\cite{Li2023,Ye2022}.

\section{Experimental implications}
The protocol is compatible with Willow-class superconducting processors ($T_1\approx70$--$100\,\mu$s). With single-shot readout fidelities of order 99 percent, the Fourier peak remains detectable after a few hundred repetitions, well within present experimental practice. A piecewise-linear landscape is programmable by flux bias; the Fourier analysis of local correlators is already standard in dynamical spectroscopy on these devices~\cite{Li2023,deAlbornoz2023}. Principal gaps open automatically at weak hopping, while a local gate realizing the bump closes a chosen non-principal gap---providing a design rule for robust versus tunable microwave stop-bands. The same signature is accessible in photonic-crystal cavity arrays~\cite{Ye2022} and, in the continuum limit, in moir\'e superlattices~\cite{Dean2013,Hunt2013}.

\section{Conclusion}
An explicit construction realizes selective closing of a non-principal gap inside the regime of the localization theory of Ref.~\cite{Kachkovskiy2026}. Large-scale numerics demonstrate a robust minimum in a weighted oscillatory correlator that directly tracks the residual eigenfunction weight controlling the gap, while the site-control remains large and featureless. The result supplies a concrete experimental test for existing quantum-simulation platforms.

\begin{acknowledgments}
L. P. was partially supported by the EPSRC grant EP/V051636/1 and the Leverhulme Trust grant RPG-2023-325. R. S. was supported by the NSF grant DMS-2306327. This work used ACES at TAMU through allocation MTH260160 from the "Advanced Cyberinfrastructure Coordination Ecosystem: Services and Support" (ACCESS)  program, which is supported by U.S. National Science Foundation grants 2138259, 2138286, 2138307, 2137603, and 2138296 (see \cite{ACCESS}). The authors acknowledge the use of LLMs for generating code for the associated numerical simulations, as well as for editorial purposes.
\end{acknowledgments}

\end{document}